\documentclass[12pt]{article}
\usepackage{amssymb,amsmath,epsfig}

\begin{document}

\title{\bf Non-vacuum Solutions of Bianchi Type $VI_0$ Universe in $f(R)$ Gravity}

\author{M. Sharif \thanks{msharif@math.pu.edu.pk} and H. Rizwana
Kausar
\thanks{rizwa\_math@yahoo.com}\\
Department of Mathematics, University of the Punjab,\\
Quaid-e-Azam Campus, Lahore-54590, Pakistan.}

\date{}
\maketitle

\begin{abstract}
In this paper, we solve the field equations in metric $f(R)$ gravity
for Bianchi type $VI_0$ spacetime and discuss evolution of the
expanding universe. We find two types of non-vacuum solutions by
taking isotropic and anisotropic fluids as the source of matter and
dark energy. The physical behavior of these solutions is analyzed
and compared in the future evolution with the help of some physical
and geometrical parameters. It is concluded that in the presence of
isotropic fluid, the model has singularity at $\tilde{t}=0$ and
represents continuously expanding shearing universe currently
entering into phantom phase. In anisotropic fluid, the model has no
initial singularity and exhibits the uniform accelerating expansion.
However, the spacetime does not achieve isotropy as
$t\rightarrow\infty$ in both of these solutions.\\
\end{abstract}

\section{Introduction}

Extended theories of gravity have become a paradigm in bringing
suitable cosmological models where a late time accelerated
expansion can be achieved. The paradigm, consisting of higher
order curvature, establishes the $f(R)$ theory of gravity. This
theory has many applications in cosmology and gravity such as
inflation, dark energy (DE), local gravity constraints,
cosmological perturbations and spherically symmetric solutions in
weak and strong gravitational backgrounds. The main motivation of
this theory comes from the fact that every unification of
fundamental interaction exhibits effective actions containing
higher order terms in the curvature invariants. This strategy was
adopted in the study of quantum field theory in curved spacetimes
\cite{1} as well as in the Lagrangian of string and Kaluza-Klein
theories \cite{2}.

Higher order terms always give an even number as an order of the
field equations. For example, $R^2$ term produces fourth order
field equations \cite{3}, term $R\Box R$ (where $\Box \equiv
\nabla^{\mu}\nabla_{\mu}$) gives sixth order field equations
\cite{4, 5}, similarly, $R\Box^2 R$ yields eights order field
equations \cite{6} and so on. Using conformal transformation, the
term with second derivative corresponds to a scalar field. For
instance, fourth order gravitational theory corresponds to
Einstein theory with one scalar field, sixth order gravity
corresponds to Einstein gravity with two scalar fields,
\textit{etc.} \cite{4,7}. In this context, it is easy to show that
$f(R)$ gravity is equivalent to scalar tensor theory as well as to
general relativity (GR) with an ideal fluid \cite{8}.

In an isotropic and homogeneous spacetimes, the Einstein field
equations give rise to the Friedmann equations which are used to
describe the evolution of the universe. However, the speedy
development in observational cosmology shows that the universe has
undergone two phases of cosmic acceleration. The first phase is
called inflation \cite{9}-\cite{12} which is believed to have
occurred earlier than matter domination \cite{13}-\cite{15}. This
accelerating phase is expected to solve horizon problems involved
in the big bang cosmology and to explain nearly flat spectrum of
temperature anisotropies observed in Cosmic Microwave Background
(CMB) radiations \cite{16}. The second accelerating phase has
occurred after the matter domination.

The first model of inflation with $R+\alpha R^2$, proposed by
Starobinsky \cite{9}, can lead to an accelerated expansion of the
universe due to $\alpha R^2$ term. The unknown component giving
rise to this late-time acceleration is called dark energy (DE)
\cite{17}-\cite{21}. The cosmic acceleration has been confirmed by
many observational sources such as Supernovae Ia (SNela)
\cite{22}-\cite{24}, Large scale structure (LSS) \cite{25,26}, CMB
\cite{27}-\cite{29} and baryon acoustic oscillations (BAO)
\cite{30,31}. The discovery of DE stimulated the idea that cosmic
acceleration might originate from some extension of GR. Dark
energy models based on $f(R)$ theory have been extensively studied
to explain the late-time acceleration.

Most of the exact solutions in this theory have been discussed in
spherical symmetry. Multam$\ddot{a}$ki and Vilja \cite{40} studied
static spherically symmetric vacuum as well as non-vacuum solutions
by taking perfect fluid \cite{47}. Caram$\hat{e}$s and Bezerra
\cite{41} found spherically symmetric vacuum solutions in higher
dimensions. Capozziello \emph{et al.} \cite{42} analyzed spherically
symmetric solutions using Noether symmetry. The structure of
relativistic stars in $f(R)$ theory has been discussed by many
authors \cite{95}-\cite{99}. In a recent paper, we have found static
spherically symmetric solution in the presence of dust matter and
discussed energy-momentum distribution for some well-known $f(R)$
models \cite{sss} using Landau-Lifshitz complex. We have also
checked the stability and constant curvature conditions of these
models.

In cylindrical symmetry, Azadi \textit{et al.} \cite{43} have
studied solutions in Weyl coordinates. They have shown that constant
curvature solutions reduce to only one member of the Tian family in
GR. Momeni \cite{44} has found that exact constant scalar curvature
solution in cylindrical symmetry is applicable to the exterior of a
string. Sharif and his collaborators \cite{45,46} have investigated
the exact solutions of plane Bianchi models for both vacuum and
non-vacuum cases. Hollestein and Lobo \cite{lobo} explored exact
solutions of the field equations coupled to non-linear
electrodynamics.

It has been observed that some large-angle anomalies appear in CMB
radiations which violate the statistical isotropy of the universe
\cite{32,33}. Plane Bianchi models (which are homogeneous but not
necessarily isotropic) seem to be the most promising explanation
of these anomalies. Jaffe \textit{et al}. \cite{34}-\cite{36}
investigated that removing a Bianchi component from the WMAP data
can account for several large-angle anomalies leaving the universe
to be isotropic. Thus the universe may have achieved a slight
anisotropic geometry in cosmological models regardless of the
inflation. Further, these models can be classified according to
whether anisotopy occurs at an early stage or at later times of
the universe. The models for the early stage can be modified in a
way to end inflation with a slight anisotropic geometry \cite{37}.
For the later class, the isotropy of the universe, achieved during
inflation, can be distorted by modifying DE \cite{38,39}.

The objective of this paper is to find non-vacuum exact solutions of
Bianchi type $VI_0$ model in the metric $f(R)$ gravity. The
organization of the paper is as follows. In section \textbf{2}, we
present the field equations and some dynamical quantities describing
the evolution of the universe. Section \textbf{3} provides solution
of the field equations in the  presence of perfect fluid. In section
\textbf{4}, we find solution for anisotropic fluid. In the last
section \textbf{5}, summary and comparison of both the solutions is
given.

\section{The Model and the Field Equations}

The line element for homogeneous and anisotropic Bianchi type
$VI_0$ spacetime is given by
\begin{equation}\label{1}
ds^{2}=dt^2-A^2(t)dx^2-e^{2\alpha x}B^2(t)dy^2-e^{-2\alpha
x}C^2(t)dz^2,
\end{equation}
where the scale factors $A,~B$ and $C$ are functions of cosmic time
$t$ only and $\alpha$ is a non-zero constant. There are two
formalisms which are applied to derive the field equations in $f(R)$
gravity. One is the standard metric formalism while another is
Palatini formalism. Most of the work in this theory has been done by
using the former formalism, where the action with $S^{(m)}$ as a
matter part, is given as follows \cite{action}
\begin{equation}\setcounter{equation}{1}\label{3.1}
S=\frac{1}{2\kappa}\int d^{4}x\sqrt{-g}f(R)+S^{(m)}.
\end{equation}
Its variation with respect to the metric tensor yields the following
set of field equations
\begin{equation}\label{3.2}
F(R) R_{\mu\nu} - \frac{1}{2}f(R)g_{\mu\nu}-\nabla_{\mu}
\nabla_{\nu}F(R)+ g_{\mu\nu} \Box F(R)= \kappa T_{\mu\nu},
\end{equation}
where $F(R)\equiv df(R)/dR$ and $f(R)$ is a function of the Ricci
scalar and describes all kinds of matter including
non-relativistic (cold) dark matter. Taking trace of the above
equation (with $\kappa=1$), we obtain $f(R)$
\begin{equation}\label{2.3}
f(R)=\frac{F(R) R+ 3\Box F(R)-T}{2}.
\end{equation}
The scalar curvature for Bianchi type $VI_0$ model is given by
\begin{equation}\label{3.R}
R=-2\left[\frac{\ddot{A}}{A}+\frac{\ddot{B}}{B}+\frac{\ddot{C}}{C}-\frac{\alpha^2}{A^2}
+\frac{\dot{A}\dot{B}}{AB}+\frac{\dot{A}
\dot{C}}{AC}+\frac{\dot{B} \dot{C}}{BC}\right].
\end{equation}

The directional Hubble parameters along $x,~y$ and $z$ directions,
the average scale factor $a$, the deceleration parameter $q$, the
expansion scalar $\Theta$, the shear scalar $\sigma$ and the mean
Hubble parameter $H$ are respectively given as
\begin{eqnarray}\label{26}
&&H_x=\frac{\dot{A}}{A},\quad H_y=\frac{\dot{B}}{B},\quad
H_z=\frac{\dot{C}}{C},\quad a=(ABC)^{1/3},
\\\label{26'}
&& q=-\frac{a\ddot{a}}{\dot{a}^2},\quad
\Theta=u_{;a}^a=\frac{\dot{A}}{A}+\frac{\dot{B}}{B}+\frac{\dot{C}}{C},\quad
\sigma^2=\frac{1}{2}\sigma_{ab}\sigma^{ab},\\\label{26''} &&
H=\frac{1}{3}({\ln}V\dot{)}=\frac{1}{3}(\frac{\dot{A}}{A}+\frac{\dot{B}}{B}
+\frac{\dot{C}}{C}).
\end{eqnarray}
Here $u^a$ is the four velocity vector and $\sigma_{ab}$ is the
shear tensor. Berman \cite{var1,var2} introduced the variation of
mean Hubble parameter as follows
\begin{equation}\label{31}
H=ka^{-n},
\end{equation}
where $k>0$ and $n\geqslant0$. Inserting this value in
Eq.(\ref{26'}), we obtain
\begin{eqnarray}\label{32}
a&=&c_1e^{kt},\qquad\quad\quad\quad q=-1  \quad \textmd{for} \quad
n=0,\\\label{33} a&=&(nkt+c_2)^{1/n},\quad q=n-1 \quad \textmd{for}
\quad n\neq0,
\end{eqnarray}
where $c_1$ and $c_2$ are positive constants of integration. For
$n<1$, these equations represent an accelerated expansion of the
universe with $a\rightarrow\infty$ as $t\rightarrow\infty$ and
supports the observations of the Type Ia supernova \cite{22,23} and
WMAP data \cite{27,28}. To examine whether expansion of the universe
is anisotropic or not, we define anisotropic expansion parameter as
\begin{equation}\label{aniso}
\Delta=\frac{1}{3}\sum_{i=1}^3(\frac{H_i-H}{H})^2.
\end{equation}
If $\Delta=0$, then the expansion of the universe is isotropic.
Further, any anisotropic model of the universe with diagonal
energy-momentum tensor approaches to isotropy if
$\Delta\rightarrow0,~V\rightarrow{+\infty}$ and $\rho>0$ as
$t\rightarrow{+\infty}$ \cite{akar,collins}.

\section{Solution with Isotropic Fluid}

In this section, we take isotropic perfect fluid whose
energy-momentum tensor is given by
\begin{equation}\setcounter{equation}{1}\label{2.02}
T_\mu^\nu=diag[\rho,-p,-p,-p],
\end{equation}
where $\rho$ is the density and $p$ is the pressure. The
corresponding field equations become
\begin{eqnarray}\label{2.05}
(\frac{\ddot{A}}{A}+\frac{\ddot{B}}{B}+\frac{\ddot{C}}{C})F
+\frac{1}{2}f(R)-(\frac{\dot{A}}{A}+\frac{\dot{B}}{B}+\frac{\dot{C}}{C})\dot{F}
=-\rho,&&\\\label{2.06}
(\frac{\ddot{A}}{A}-\frac{2\alpha^2}{A^2}+\frac{\dot{A}\dot{B}}{AB}+\frac{\dot{A}
\dot{C}}{AC})F
+\frac{1}{2}f(R)-\ddot{F}-(\frac{\dot{B}}{B}+\frac{\dot{C}}{C})\dot{F}
=p,&&\\\label{2.07}
(\frac{\ddot{B}}{B}+\frac{\dot{A}\dot{B}}{AB}+\frac{\dot{B}
\dot{C}}{BC})F
+\frac{1}{2}f(R)-\ddot{F}-(\frac{\dot{A}}{A}+\frac{\dot{C}}{C})\dot{F}
=p,&&\\\label{2.08}
(\frac{\ddot{C}}{C}+\frac{\dot{A}\dot{C}}{AC}+\frac{\dot{B}\dot{C}}{BC})F
+\frac{1}{2}f(R)-\ddot{F}-(\frac{\dot{A}}{A}+\frac{\dot{B}}{B})\dot{F}
=p,&&\\\label{2.09}
\alpha(\frac{\dot{C}}{C}-\frac{\dot{B}}{B})F=0.&&
\end{eqnarray}
The solution of Eq.(\ref{2.09}) yields
\begin{equation}\label{3.010}
C=c_3B,
\end{equation}
where $c_3>0$ is another constant of integration. Without any loss
of generality, we take $c_3=1$ for the sake of simplicity. Using
this value of $C$ in the above equations, we obtain
\begin{eqnarray}\label{2.012}
(\frac{\ddot{A}}{A}+\frac{\ddot{2B}}{B})F
+\frac{1}{2}f(R)-(\frac{\dot{A}}{A}+\frac{2\dot{B}}{B})\dot{F}
&=&-\rho,\\\label{2.013}
(\frac{\ddot{A}}{A}-\frac{2\alpha^2}{A^2}+\frac{2\dot{A}\dot{B}}{AB})F
+\frac{1}{2}f(R)-\ddot{F}-\frac{2\dot{B}}{B}\dot{F}&=&p,
\\\label{2.014}
(\frac{\ddot{B}}{B}+\frac{\dot{A}\dot{B}}{AB}+\frac{\dot{B}^2}{B^2})F+\frac{1}{2}f(R)
-\ddot{F}-(\frac{\dot{A}}{A}+\frac{\dot{B}}{B})\dot{F}&=&p.
\end{eqnarray}
Subtracting Eq.(\ref{2.013}) from (\ref{2.012}), Eq.(\ref{2.014})
from (\ref{2.012}) respectively and dividing the resulting equations
by $F$, we have
\begin{eqnarray}\label{2.015}
\frac{-\ddot{B}}{B}-\frac{2\alpha^2}{A^2}+\frac{2\dot{A}\dot{B}}{AB}
-\frac{\ddot{F}}{F}+\frac{\dot{A}}{A}\frac{\dot{F}}{F}&=&\frac{\rho+p}{F},
\\\label{2.016}
-\frac{\ddot{A}}{A}-\frac{\ddot{B}}{B}+\frac{\dot{A}\dot{B}}{AB}+\frac{\dot{B}^2}{B^2}
-\frac{\ddot{F}}{F}+\frac{\dot{B}}{B}\frac{\dot{F}}{F}&=&\frac{\rho+p}{F}.
\end{eqnarray}

We assume that the scalar expansion is proportional to the shear
scalar \cite{singh}-\cite{adhav}, i.e., $\Theta\propto\sigma$,
which leads to a relation between the metric functions as follows
\begin{equation}\label{3.017}
B=A^n,
\end{equation}
where $n\neq1$ is a positive constant. Using this relation in
Eqs.(\ref{2.015}) and (\ref{2.016}), it follows that
\begin{eqnarray}\label{3.018}
2n(2-n)\frac{\dot{A}^2}{A^2}-2n\frac{\ddot{A}}{A}-\frac{2\alpha^2}{A^2}
-\frac{\ddot{F}}{F}+\frac{\dot{A}}{A}\frac{\dot{F}}{F}&=&\frac{\rho+p}{F},\\\label{3.019}
2n\frac{\dot{A}^2}{A^2}-(n+1)\frac{\ddot{A}}{A}-\frac{\ddot{F}}{F}
+n\frac{\dot{A}}{A}\frac{\dot{F}}{F}&=&\frac{\rho+p}{F}.
\end{eqnarray}
Subtraction of Eq.(\ref{3.019}) from (\ref{3.018}) yields
\begin{equation}\label{3.020}
\frac{\ddot{A}}{A}+2n\frac{\dot{A}^2}{A^2}
+\frac{\dot{A}\dot{F}}{AF}-\frac{2\alpha^2}{(1-n)A^2}=0.
\end{equation}
We solve this equation using power law relation between $F$ and
$a$ \cite{45},
\begin{equation}\label{3.021}
F=la^m,
\end{equation}
where $l$ is the constant of proportionality and $m$ is any real
number. Substituting the value of $a$, $F$ turns out to be
\begin{equation}\label{23}
F=lA^{(\frac{2n+1}{3})m}.
\end{equation}
Inserting this value in Eq.(\ref{3.020}), it follows that
\begin{equation}\label{3.024}
\frac{\ddot{A}}{A}+\frac{6n+(2n+1)m}{3}\frac{\dot{A}^2}{A^2}-\frac{2\alpha^2}{(1-n)A^2}=0.
\end{equation}
Integrating this equation, we obtain
\begin{equation}\label{3.026}
\dot{A}=\sqrt{\frac{6\alpha^2}{(1-n)(m+6n+2mn)}+\frac{c_4}{A^{2(m+6n+2mn)/3}}},
\end{equation}
where $c_4$ is another integration constant. With the help of
Eqs.(\ref{3.010}) and (\ref{3.017}), the line element (\ref{1})
reduces to
\begin{eqnarray}\nonumber
ds^{2}&=&\left[\frac{(1-n)(m+6n+2mn)A^{2(m+6n+2mn)/3}}{6\alpha^2A^{2(m+6n+2mn)/3}
+c_4(m+6n+2mn)}\right]dA^2-A^2dx^2\\\label{27} &-& e^{2\alpha
x}A^{2n}dy^2-c_3^2A^{2n}e^{-2\alpha x}dz^2.
\end{eqnarray}
Taking the coordinate transformation ($A=\tilde{t}$),
Eq.(\ref{27}) becomes
\begin{eqnarray}\nonumber
ds^{2}&=&\left[\frac{(1-n)(m+6n+2mn)\tilde{t}^{2(m+6n+2mn)/3}}{6\alpha^2
\tilde{t}^{2(m+6n+2mn)/3}+c_4(m+6n+2mn)}\right]d\tilde{t}^2-\tilde{t}^2dx^2\\\label{28}
&-&e^{2\alpha x}\tilde{t}^{2n}dy^2-c_3^2\tilde{t}^{2n}e^{-2\alpha
x}dz^2.
\end{eqnarray}

Using Eq.(\ref{23}) in  Eqs.(\ref{3.018}) and (\ref{3.019}), we
obtain
\begin{eqnarray}\nonumber
&&[2n(2-n)+\frac{2(2n+1)m}{3}-\frac{(2n+1)^2m^2}{9}]\frac{\dot{A}^2}{A^2}\\\label{2.018}
&&-[2n+\frac{(2n+1)m}{3}] \frac{\ddot{A}}{A}-\frac{2\alpha^2}{A^2}
=\frac{\rho+p}{lA^{\frac{(2n+1)m}{3}}},\\\nonumber
&&[2n+(1-n)\frac{(2n+1)m}{3}-\frac{(2n+1)^2m^2}{9}]\frac{\dot{A}^2}{A^2}\\\label{2.019}
&&-[(n+1)+\frac{(2n+1)m}{3}]
\frac{\ddot{A}}{A}=\frac{\rho+p}{lA^{\frac{(2n+1)m}{3}}}.
\end{eqnarray}
Since the right hand sides of these equations are same with two
unknown functions $\rho$ and $p$, thus we get a dependent solution.
Using equation of state (EoS) $p=\omega\rho$, where $\omega$ is the
EoS parameter and adding Eqs.(\ref{2.018}) and (\ref{2.019}), the
energy density is found to be
\begin{eqnarray}\nonumber
\rho&=&\frac{l\tilde{t}^{\frac{(2n+1)m}{3}-2}}{2(1+\omega)}
[\{\frac{-6\alpha^2}{(n-1)(m+6n+2mn)}+\frac{c_4}{\tilde{t}^{2(m+6n+2mn)/3}}\}\\\nonumber
&\times&\{6n^2+m(-2n^2+\frac{25n}{12}+\frac{5}{3})+\frac{8}{9}(2n+1)^2
m^2\}\\
&+&4\alpha^2\{\frac{n+1}{n-1}+(\frac{2n+1}{1-n})\frac{m}{3}\}].
\end{eqnarray}
The positive values of $\omega$ characterize different kinds of
fluids while in order to explain DE of the current universe with
the help of Bianchi models, one can use $\omega$ with negative
values. When $w=-1$, the universe passes through $\Lambda$CDM
epoch. If $w<-1$, then we live in the phantom-dominated universe
and for $w>-1$, the quintessence dark era occurs. Here we analyze
the behavior of energy density in phantom and quintessence regions
for different positive values of $m$ as shown in Figure \textbf{1}
while $\omega=-1$ yields singular expression. It can be observed
that our solution favors only the phantom region because $\rho$ is
negative in the other region which is not feasible. Thus the
energy density will increase and the universe enters into the
phantom phase. It is worthwhile to mention here that observational
analysis of a recent supernova strongly supports $w<-1$ being EoS
parameter for phantom DE \cite{A}-\cite{S}.
\begin{figure}\center
\begin{tabular}{cccc}
& Phantom region & Quintessence region\\ &
\epsfig{file=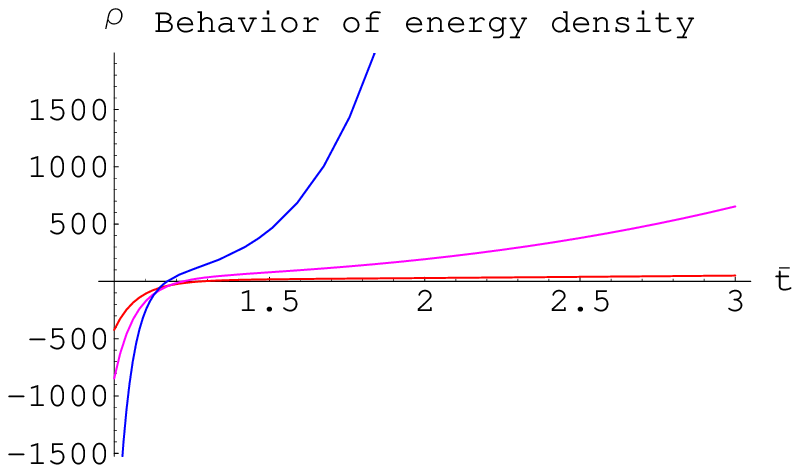,width=0.4\linewidth} &
\epsfig{file=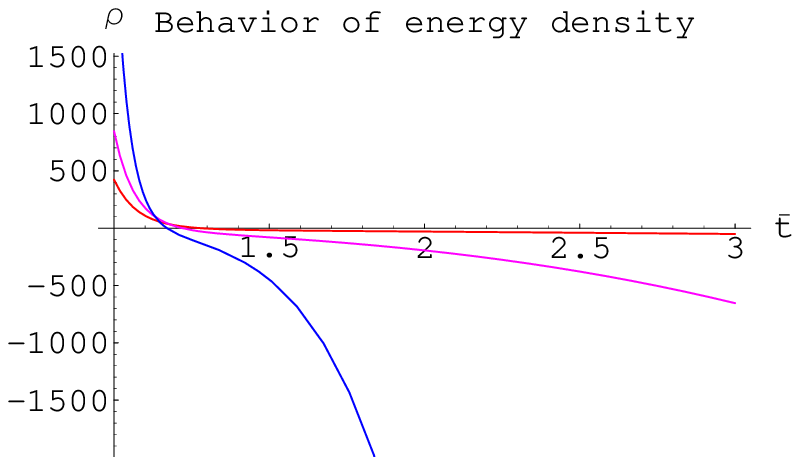,width=0.4\linewidth} \\
\end{tabular}
\caption{Behavior of energy density versus cosmic time. The three
colored lines are sketched for $m=2,~m=3$ and $m=4$
respectively.}\center
\end{figure}

For this model, the physical quantities will become
\begin{eqnarray}\nonumber
H&=&\frac{(2n+1)}{3\tilde{t}^{m+6n+2mn+1}}\sqrt{-2\alpha^2
\tilde{t}^{2(m+6n+2mn)}+c_4(n-1)(m+6n+2mn)},\\\label{3.030}
V&=&\tilde{t}^{2n+1}.
\end{eqnarray}
The expansion scalar $\Theta$ and the shear scalar $\sigma$ become
\begin{eqnarray}\label{3.032}
\Theta=\frac{(2n+1)(-6\alpha^2
\tilde{t}^{2(m+6n+2mn)/3}+c_4(n-1)(m+6n+2mn))^{1/2}}
{\tilde{t}^{\frac{m+6n+2mn}{3}+1}},&&\\\label{3.033}
\sigma=\frac{(1-n)(-6\alpha^2
\tilde{t}^{2(m+6n+2mn)}+c_4(n-1)(m+6n+2mn))^{1/2}}{\sqrt{3}
~\tilde{t}^{\frac{m+6n+2mn}{3}+1}}.&&
\end{eqnarray}
We see that the scale factors and volume of the universe are zero at
initial epoch which shows that the model has point type singularity
while they continue to increase with time. The Hubble parameter and
the expansion scalar indicate that expansion rate was rapid at
initial times of the big bang but it slows down with the passage of
time and tends to zero as $\tilde{t}\rightarrow\infty$. The ratio
$\frac{\sigma}{\Theta}$ indicates that the universe does not achieve
isotropy and hence model represents continuously expanding, shearing
universe from the start of the big bang.

The scalar curvature and $f(R)$ function turn out to be
\begin{eqnarray}\nonumber
R&=&\frac{2\alpha^2(5n+1)}{\tilde{t}^2(n-1)}+2(5n^2+6n+(2n+1)^2m)\\\label{3.035}
&\times&\frac{-2\alpha^2\tilde{t}^{2(m+6n+2mn)}+c_4(n-1)(m+6n+2mn)}
{(n-1)(m+6n+2mn)\tilde{t}^{2(m+6n+2mn+1)}},\\\nonumber
f(R)&=&l\tilde{t}^{(2n+1)/3}R+3(n+1)\\\label{3.036} &\times&
[\frac{-2\alpha^2}{(n-1)(m+6n+2mn)\tilde{t}^2}+\frac{c_4}{\tilde{t}^{2(m+6n+2mn+1)}}]^{1/2}.
\end{eqnarray}
We know that if $f(R)$ is replaced by $R+\Lambda$, then $f(R)$
theory corresponds to GR, where $\Lambda$ is interpreted as the
energy density of the vacuum \cite{48}-\cite{51} causing expansion
in the universe. Thus $f(R)$ may be used to explain the present
cosmological expansion. We note that the above function shows
initial time singularity and continuously expanding to infinity.

\section{Solution with Anisotropic Fluid}

Here, we obtain solution of the field equations for anisotropic
fluid
\begin{eqnarray}\setcounter{equation}{1}\label{2}
T_{\mu}^{\nu}&=&diag[\rho,-p_{x},-p_{y},-p_{z}]=diag[1,-{\omega}_{x}
,-{\omega}_{y},-{\omega}_{z}]\rho,
\end{eqnarray}
where $p_x,~p_y$ and $p_z$ are pressures and $\omega_x,~\omega_y$
and $\omega_z$ are directional EoS parameters on $x,~y$ and $z$ axes
respectively. We take
\begin{equation*}
{\omega}_x={\omega}+{\delta}, \quad {\omega}_y={\omega} \quad
\textmd{and} \quad {\omega}_{z}={\omega}+{\gamma},
\end{equation*}
where $\omega$ is the deviation-free EoS parameter and $\delta$ and
$\gamma$ (called skewness parameters) are deviations from $\omega$
on $x$ and $z$ axes respectively. The energy-momentum tensor takes
the form
\begin{equation}\label{12}
T_{\mu}^{\nu}=diag[1,-({\omega}+{\delta}),-{\omega},-({\omega}+{\gamma})]\rho.
\end{equation}
Notice that $\omega,~\delta$ and $\gamma$ are functions of cosmic
time $t$. The field equations will become
\begin{eqnarray}\label{3.5}
(\frac{\ddot{A}}{A}+\frac{\ddot{B}}{B}+\frac{\ddot{C}}{C})F
+\frac{1}{2}f(R)-(\frac{\dot{A}}{A}+\frac{\dot{B}}{B}+\frac{\dot{C}}{C})\dot{F}
=-\rho,&&\\\label{3.6}
(\frac{\ddot{A}}{A}-\frac{2\alpha^2}{A^2}+\frac{\dot{A}\dot{B}}{AB}+\frac{\dot{A}
\dot{C}}{AC})F
+\frac{1}{2}f(R)-\ddot{F}-(\frac{\dot{B}}{B}+\frac{\dot{C}}{C})\dot{F}
=(\omega+\delta)\rho,&&\\\label{3.7}
(\frac{\ddot{B}}{B}+\frac{\dot{A}\dot{B}}{AB}+\frac{\dot{B}
\dot{C}}{BC})F
+\frac{1}{2}f(R)-\ddot{F}-(\frac{\dot{A}}{A}+\frac{\dot{C}}{C})\dot{F}
=\omega\rho,&&\\\label{3.8}
(\frac{\ddot{C}}{C}+\frac{\dot{A}\dot{C}}{AC}+\frac{\dot{B}\dot{C}}{BC})F
+\frac{1}{2}f(R)-\ddot{F}-(\frac{\dot{A}}{A}+\frac{\dot{B}}{B})\dot{F}
=(\omega+\gamma)\rho,&&\\\label{3.9}
\alpha(\frac{\dot{C}}{C}-\frac{\dot{B}}{B})F=0.&&
\end{eqnarray}

Using solution of Eq.(\ref{3.9}) in the above system and then
subtracting Eq.(\ref{3.7}) from (\ref{3.8}), we obtain $\gamma=0$.
This indicates that the directional EoS parameters
$\omega_y,~\omega_z$ along $y$ and $z$ axes become equal and hence
also pressure. Consequently, the field equations turn out to be
\begin{eqnarray}\label{3.12}
(\frac{\ddot{A}}{A}+\frac{\ddot{2B}}{B})F
+\frac{1}{2}f(R)-(\frac{\dot{A}}{A}+\frac{2\dot{B}}{B})\dot{F}
&=&-\rho,\\\label{3.13}
(\frac{\ddot{A}}{A}-\frac{2\alpha^2}{A^2}+\frac{2\dot{A}\dot{B}}{AB})F
+\frac{1}{2}f(R)-\ddot{F}-\frac{2\dot{B}}{B}\dot{F}&=&(\omega+\delta)\rho,
\\\label{3.14}
(\frac{\ddot{B}}{B}+\frac{\dot{A}\dot{B}}{AB}+\frac{\dot{B}^2}{B^2})F+\frac{1}{2}f(R)
-\ddot{F}-(\frac{\dot{A}}{A}+\frac{\dot{B}}{B})\dot{F}&=&\omega\rho.
\end{eqnarray}
Subtracting Eq.(\ref{3.13}) from (\ref{3.14}) and integrating the
resulting equation, we have
\begin{equation}\label{3.16}
H_x-H_y=\frac{c_5}{VF}+
\frac{1}{VF}\int{(\frac{\alpha^2F}{A^2}+\delta\rho)Vdt},
\end{equation}
where $c_5$ is another positive constant of integration. Using
Eqs.(\ref{aniso}) and (\ref{3.16}), it follows that
\begin{equation}\label{3.17}
\Delta=\frac{2}{9H^2}[c_5+\int{(\frac{\alpha^2F}{A^2}+\delta\rho)}Vdt]^2V^{-2}F^{-2}.
\end{equation}
If we take $\delta=0$ and $F(R)=1$, the anisotropy parameter of
expansion reduces to GR for an isotropic fluid. To avoid the
integral term in Eq.(\ref{3.17}), we can choose $\delta$ such that
\begin{equation}\label{3.19}
\delta=-\frac{2\alpha^2F}{\rho A^2}.
\end{equation}
The energy-momentum tensor, the anisotropy parameter and
Eq.(\ref{3.16}) will take the form
\begin{eqnarray}\label{3.20}
T_{\mu}^{\nu}&=&diag[1,-\omega+\frac{2\alpha^2F}{\rho
A^2},-{\omega},-\omega]\rho,\\\label{3.21}
\Delta&=&\frac{2}{9}\frac{c_5^2}{H^2}V^{-2}F^{-2},\\\label{3.22}
H_x-H_y&=&\frac{c_5}{VF}.
\end{eqnarray}
Using Eq.(\ref{3.19}) in (\ref{3.13}), we obtain the following set
of field equation
\begin{eqnarray}\label{4.1}
(\frac{\ddot{A}}{A}+\frac{2\ddot{B}}{B})F
+\frac{1}{2}f(R)-(\frac{\dot{A}}{A}+\frac{2\dot{B}}{B})\dot{F}
&=&-\rho,\\\label{4.2}
(\frac{\ddot{A}}{A}+\frac{2\dot{A}\dot{B}}{AB})F
+\frac{1}{2}f(R)-\ddot{F}-\frac{2\dot{B}}{B}\dot{F}&=& \omega\rho,
\\\label{4.3}
(\frac{\ddot{B}}{B}+\frac{\dot{A}\dot{B}}{AB}+\frac{\dot{B}^2}{B^2})F
+\frac{1}{2}f(R)-\ddot{F}-(\frac{2\dot{A}}{A}+\frac{2\dot{B}}{B})\dot{F}&=&\omega\rho.
\end{eqnarray}

We can solve this system of equations by using exponential and power
law volumetric expansions corresponding to Eqs.(\ref{32}) and
(\ref{33}) respectively. The exponential expansion (also known as de
Sitter expansion) has been discussed in detail \cite{dsit}. Using
Eq.(\ref{32}), the volume of the universe and $F$ can be written as
follows
\begin{eqnarray}\label{4.4}
V&=&c_1^3e^{3kt},\\ \label{4.4'} F&=&lc_1^m e^{mkt}.
\end{eqnarray}
Inserting Eqs.(\ref{4.4}) and (\ref{4.4'}) in (\ref{3.22}), we have
\begin{equation}\label{4.5}
H_x-H_y=\frac{c_5 e^{-(3+m)kt}}{lc_1^{3+m}}.
\end{equation}
Solving the system of equations (\ref{4.1})-(\ref{4.3}) along with
Eq.(\ref{4.5}), we obtain the scale factors as follows
\begin{eqnarray}\label{4.6}
A&=&c_6e^{kt-\frac{2}{3}\frac{c_5}{l(3+m)k
c_1^{3+m}}e^{-(3+m)kt}},\\\label{4.7}
B&=&C=c_7e^{kt+\frac{1}{3}\frac{c_5}{l(3+m)k
c_1^{3+m}}e^{-(3+m)kt}},
\end{eqnarray}
where $c_6$ and $c_7$ are positive constants of integration.
\begin{figure}
\begin{center}
\begin{tabular}{cccc}
& a & b \\
& \epsfig{file=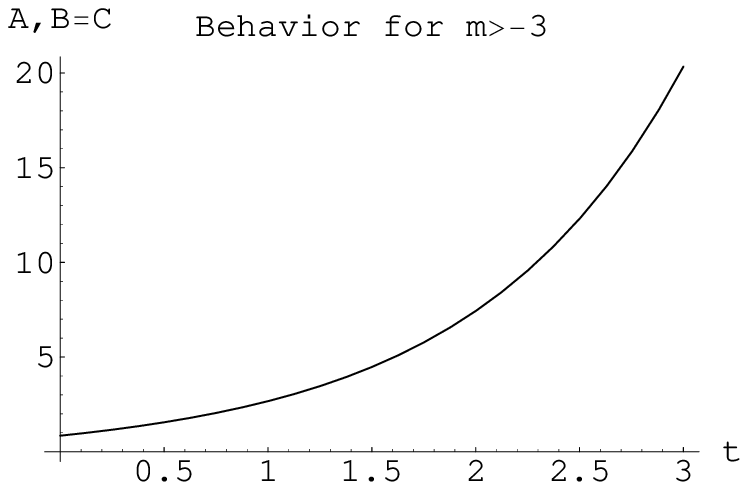,width=0.35\linewidth} &
\epsfig{file=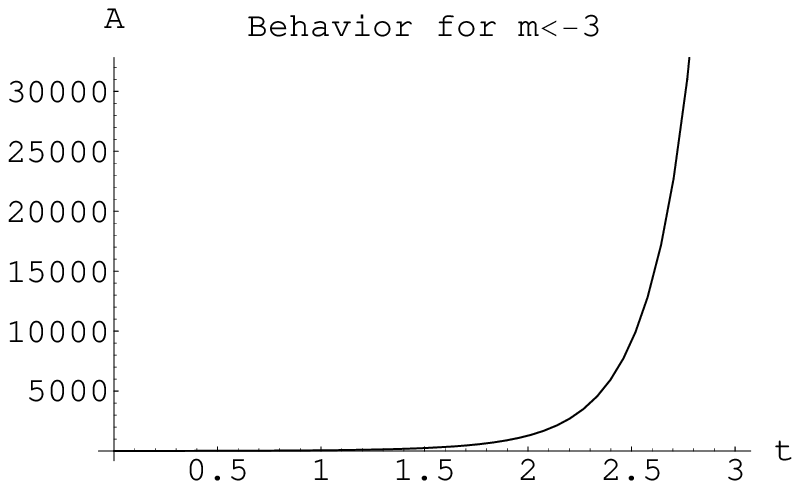,width=0.35\linewidth} \\
\end{tabular}
\end{center}
\end{figure}
\begin{figure}
\begin{center}
\begin{tabular}{cc}
c\\
\epsfig{file=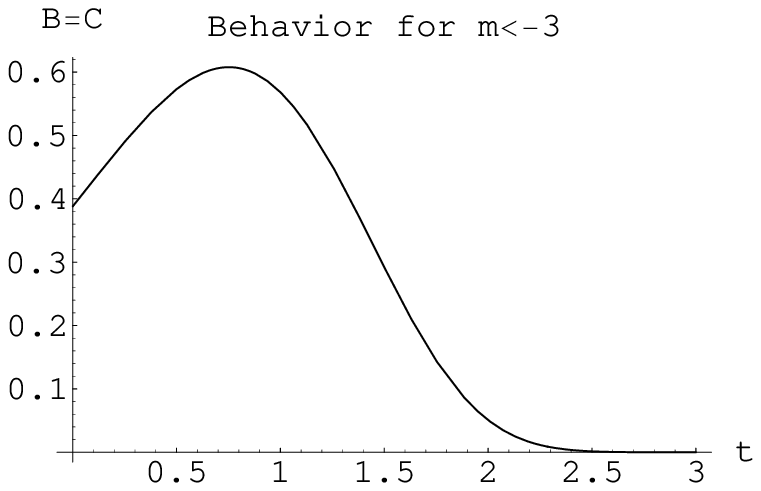,width=0.35\linewidth}\\
\end{tabular}
\caption{Behavior of scale factors versus time}
\end{center}
\end{figure}
When $m>-3$, the scale factors admit constant values at initial time
afterwards they start increasing with cosmic time without any type
of initial singularity and finally diverges to $\infty$ as
$t\rightarrow\infty$ (Figure \textbf{2a}). This shows that at the
initial epoch, the universe starts with some constant volume and
expands exponentially approaching to infinite volume. However, for
$m<-3$, the scale factor $A$ increases rapidly with time (Figure
\textbf{2b}) while $B$ first increases for a finite value of time
and then decreases approaching to zero at later times (Figure
\textbf{2c}). Moreover, the expansion scalar is constant showing
that the universe is expanding uniformly from $t=0$ to $t=\infty$.

The directional and mean Hubble parameters become
\begin{eqnarray}\label{4.13}
H_x&=&k+\frac{2c_5}{3lc_1^{3+m}}e^{-(3+m)kt}, \quad
H_y=H_z=k-\frac{c_5}{3lc_1^{3+m}}e^{-(3+m)kt},\nonumber\\
H&=&k.
\end{eqnarray}
The anisotropy parameters turns out to be
\begin{equation}
\Delta=\frac{2}{9}\frac{c_5^2}{k^2l^2c_1^{2(3+m)k}}e^{-2(m+3)kt}.
\end{equation}
We see that the mean Hubble parameter is constant whereas others are
time dependent. As time approaches from zero to infinity (for
$m>-3$), the directional Hubble parameters are reduced to the mean
value. For $m<-3$, parameter along $x$-axis will increase from the
mean Hubble parameter by twice a constant factor (in the coefficient
of exponential function) whereas parameters along $y$ and $z$-axes
decrease by the same factor. Notice that at $t=0$, the anisotropy
parameter measures a constant value while it vanishes for $m>-3$ at
infinite time of the universe. This indicates that the universe
expands isotropically at later times without taking any effect from
anisotropy of the fluid. However, for $m<-3$, the anisotropy in the
expansion will increase with time.

Using values of the scale factors in Eq.(\ref{3.12}), we obtain the
following energy density of anisotropic dark fluid
\begin{equation}\label{4.15}
\rho=3l(m-1)k^2c_1^m
e^{mkt}-\frac{1}{2}f(R)+\frac{2c_5^2}{3lc_1^{6+m}}e^{-(6+m)kt}
\end{equation}
which involves $f(R)$ function. Thus if we interpret this function
as a simplest model corresponding to $\Lambda$CDM, we may observe
that at the initial epoch, the energy density is proportional to a
constant vacuum energy density $\Lambda$ only. However, with the
passing time, it acquires dynamical terms describing matter
densities of fluids filling the universe. These terms will expand to
infinity in the future evolution of the universe. This indicates
that the energy density is increasing in the future due to creation
of some phantom DE and decay of some other components of energy in
the universe, like, cold dark matter. Consequently, expansion of the
universe is accelerated forever.

Inserting the values of $A,~F,~\rho$ in Eq.(\ref{3.19}), we obtain
skewness parameter $\delta$
\begin{equation}\label{4.17}
\delta=\frac{-12\alpha^2l^2c_1^{2(3+m)}e^{mkt-2kt+\frac{4}{3}\frac{c_5}{l(3+m)k
c_1^{3+m}}e^{-(3+m)kt}}}{c_6^2[18l^2k^2(m-1)c_1^{4+m}e^{mkt}
-4c_5^2c_1^{-(2+m)}e^{-(6+m)kt}-3lc_1^4f(R)]}.
\end{equation}
Using this value of $\delta$ alongwith scale factors and $\rho$ in
Eqs.(\ref{4.1})-(\ref{4.3}) and solving them simultaneously, the
anisotropic EoS parameter $\omega$ will become
\begin{equation}\label{4.16}
\omega=\frac{6(1-m)c_1^{3+m}l^2k^2e^{mkt}
-8c_5lmke^{-3kt}+6lc_1^3f(R)}{18l^2k^2(m-1)c_1^{4+m}e^{mkt}
-4c_5^2c_1^{-(2+m)}e^{-(6+m)kt}-3lc_1^4f(R)}.
\end{equation}
Since both the parameters contain $\rho$ and hence $f(R)$ function,
it may not be possible to discuss their behavior independently in
the future evolution. The scalar curvature for exponential solution
becomes
\begin{equation}\label{4.18}
R=-10k^2-\frac{2}{3}\frac{c_5^2}{l^2c_1^{2(3+m)}}e^{-2(3+m)kt}+
\frac{2\alpha^2}{c_6^2} e^{-2kt+\frac{4}{3}\frac{c_5}{l(3+m)k
c_1^{3+m}}e^{-(3+m)kt}}.
\end{equation}
This procedure can be extended to the solution of the field
equations for power-law volumetric expansion.

\section{Outlook}

This paper is devoted to study the exact solutions of the Bianchi
type $VI_0$ universe in the metric $f(R)$ gravity and to discuss the
recent cosmic acceleration. Two types of non-vacuum solutions are
found corresponding to isotropic and anisotropic fluids
respectively. The physical behavior of these solutions is discussed
at early and late times of the universe, which can be summarized as
follows:

\begin{itemize}
\item The scale factors in the isotropic fluid case are zero at $\tilde{t}=0$ which
shows that the spacetime exhibits point type singularity and
continues to expand till $\tilde{t}\rightarrow\infty$. In
anisotropic case, the scale factors admit constant value at initial
epoch and then start increasing with cosmic time approaching to very
large values as $t\rightarrow\infty$. Thus, the spacetime does not
show any type of initial singularity in this case.

\item The model starts with physical parameters such that all being
infinite at early times and approaches to zero at later times in the
perfect fluid case. In the anisotropic case, the expansion scalar
and the mean Hubble parameter are constant indicating homogenous
expansion of the universe. The volume of the universe will increase
with time due to expansion for both the solutions.

\item The matter density and pressure of the fluid are related by
the EoS, where EoS parameter is used to characterize the DE into
different expansion histories. It is found that in isotropic
solution, the DE has large negative pressure with $\omega<-1$ which
shows that the universe passes through phantom region. It is
mentioned here that our conclusions support the observational
evidence of a recent supernova data \cite{A}-\cite{S}. However, the
energy density of the anisotropic fluids involves $f(R)$ function
whose interpretation is given in the context of $\Lambda CDM$. In
this case, the EoS parameter turns out to be time dependent.

\item In the anisotropic solution, the deviation from the isotropy along
$z$-axis is zero which shows that pressure of DE along $y$ and $z$
axes are same. The skewness parameter along $x$-axis does not vanish
even at the later times. However, for $m>-3$, the anisotropy
parameter of the expansion becomes zero at infinite times showing
that the universe expands isotropically regardless of the anisotropy
in the fluid. It is mentioned here that in GR, the anisotropic fluid
as well as the model attain isotropy in the future evolution of the
universe \cite{akar,zubair} but in $f(R)$ theory, the model and the
fluid remain anisotropic. Further, in the isotropic fluid case, the
universe does not achieve isotropy and the model expands
continuously with non-zero shear scalar. It is remarked here that
the above mentioned aspect of the model is compatible with existing
results of GR \cite{singh,adhav}.
\end{itemize}

\vspace{0.25cm}

{\bf Acknowledgment}

\vspace{0.25cm}

We would like to thank the Higher Education Commission, Islamabad,
Pakistan for its financial support through the {\it Indigenous
Ph.D. 5000 Fellowship Program Batch-III}.

\end{document}